\documentclass[pre,twocolumn]{revtex4-2}
\usepackage{amsmath,amssymb}
\usepackage{graphicx}
\usepackage{color}
\usepackage[a-1b]{pdfx}

\begin{document}
\title{Glauber dynamics of the Sherrington--Kirkpatrick spin-glass model: a dynamical cavity method}
\author{Grzegorz Szamel}

\affiliation{Department of Chemistry,
  Colorado State University, Fort Collins, CO 80523, USA}

\begin{abstract}
  We present a closed theory, exact in the $N\to\infty$ limit,
  of the equilibrium spin correlations of the Sherrington--Kirkpatrick spin-glass model evolving with Glauber
  dynamics, above and at the transition temperature $T_c$. The derivation uses a dynamic cavity approach: 
  in the evolution equation for the selected spin the other spins are represented by a fluctuating
  field. The influence of the selected spin on this field is included at the linear response level;
  it is given as a sum of the spin's past flips weighted by the correlations of the
  unperturbed fluctuating field. The self-consistency condition relates the time-dependent correlations of the
  selected spin and those of the fluctuating field. The theory predicts a $(T-T_c)^{-2}$ divergence of the relaxation
  time upon approaching the transition and an $At^{-1/2}$ decay of spin correlations at $T_c$. A numerical solution
  of the theory gives the amplitude, $A=0.964\pm 0.007$. The theoretical predictions are in excellent agreement
  with Glauber dynamics computer simulations.
\end{abstract}

\maketitle

\section{Introduction}
The Sherrington--Kirkpatrick (SK) \cite{SK} spin-glass model is the prototypical model of a disordered system.
Its fully connected structure allows for a complete analysis of its properties. The analysis is relatively
simple in the paramagnetic phase \cite{SK,KS,BY} but becomes highly nontrivial below the de Almeida--Thouless
\cite{AT} line, where replica symmetry breaks down. The full replica-symmetry-breaking solution
\cite{Parisi1,Parisi2} resulted in a plethora of further studies and insights, culminating in the famous
``Spin Glass Theory and Beyond'' monograph \cite{SGTandB}. It should be noted, however, that
while this monograph's presentation of the static properties focuses on Ising spins, its discussion of the
dynamics uses the model of continuous or ``soft'' spin variables, with a single-spin potential that favors
$\pm 1$ values. This approach requires a perturbative analysis in terms of the coupling constant
of the single-spin potential \cite{SZ} and makes quantitative comparison with computer simulations, which typically
use Glauber dynamics of Ising spins \cite{Glauber}, difficult. 

The phenomenology of the dynamics at zero external field and at temperatures above and at the
spin-glass transition, $T\ge T_c$, is relatively simple. Upon approaching the transition the dynamics
slows down and the relaxation time diverges. At $T_c$, sample-averaged spin correlations $C(t)$ exhibit
critical decay; asymptotically $C(t) \simeq A t^{-1/2}$. While the qualitative prediction of the critical
decay goes back to Kirkpatrick and Sherrington \cite{KS}, so far no theoretical approach has been able to predict
its exact amplitude. This is on the one hand surprising in view of the simplicity of the model and, on
the other hand, problematic since the lack of a quantitative theory impacts the analysis of related
spin glass and neural network models. 

Glauber dynamics of the Ising spin SK model was previously studied by Sommers \cite{Sommers}.
His final analysis relied upon a perturbative expansion in $(\beta J)^2$, where $\beta=1/T$
(we set $k_B=1$) and $J$ quantifies the strength of
the random coupling constants. The resulting formula was quantitatively incorrect at
short times and significantly underestimated the amplitude of the critical decay \cite{Szamelsg1}. 
Our own earlier attempts \cite{Szamelsg1,Szamelsg2} resulted in expressions for spin correlations that are
exact at short times but, as we show below, underestimate the amplitude of the critical decay.
The path forward
from these theories to the exact description at all $T\ge T_c$ was unclear.

We should note that zero-temperature
parallel dynamics of the Ising spin SK model was previously analyzed using a generating
functional approach \cite{EO,EBKZ}. However, these works analyzed zero-temperature quenches, and
parallel dynamics does not sample the Gibbs distribution.
Also, Glauber dynamics of the Ising spin SK model with independent (fully asymmetric)
coupling constants was solved analytically \cite{CS} and then analyzed rigorously
\cite{Grunwald}, but the asymmetric case is different from (and simpler than) the original SK model
with symmetric couplings. 

Here we present a new approach to Glauber dynamics of the SK model at $T\ge T_c$. We use a dynamic
cavity approach: the exact evolution law for the selected spin is kept; the remaining spins, \textit{i.e.}
the environment, influence the selected spin through a fluctuating local field. The local field is in turn influenced
by the selected spin; for the SK model in the thermodynamic limit, this influence is given exactly
by linear response. The correlations of the unperturbed part of the local field can be expressed in terms of the 
time-dependent spin correlation function. Finally, the selected spin time-dependent correlations can
be evaluated from its evolution law. The self-consistency condition, which imposes equality of the
selected spin correlations and environment spin correlations, closes the theory.
Similarly to the previous analysis of parallel dynamics \cite{EO}, our final evolution law for
the single selected spin is expressed in terms of a stochastic process whose statistics are set by the 
time-dependent spin correlations. In contrast to spherical spin-glass models, where one can derive closed,
explicit equations for the correlation and response functions \cite{CK,CC},
here there is no closed, explicit equation of motion for spin correlations. 

The resulting theory reproduces the well-known phenomenology: the relaxation time of spin correlations diverges
as $(T-T_c)^{-2}$ when $T$ approaches $T_c$. At $T_c$ and at long times, spin correlations exhibit
a critical decay $At^{-1/2}$. The theory allows for a numerical evaluation of the amplitude, $A=0.964\pm 0.007$.
The theoretical predictions are compared with Glauber dynamics computer simulations. In particular,
at $T_c$, simulations of $N=256$--$16384$
spins were run and the results were extrapolated to the $N\to\infty$ limit. The theoretical
predictions are in excellent agreement with the extrapolated results.  

\section{Model and evolution law for the selected spin} 
We consider the SK model with Ising spins $\sigma_i=\pm 1$,
\begin{equation}\label{eq:H}
H=-\sum_{i<j}J_{ij}\sigma_i\sigma_j ,
\end{equation}
with independent Gaussian coupling constants $J_{ij}=J_{ji}$ of mean $0$ and variance $J^2/N$. 
Each spin has a Poisson clock of rate 1 and is reset by the heat-bath rule at its clock rings, \textit{i.e.} flips with rate
$w_i=\frac{1}{2}[1-\sigma_i\tanh\beta h_i]$ where 
the local field reads $h_i=\sum_j J_{ij}\sigma_j$. The rate satisfies the detailed balance condition
with respect to $\exp\left(-\beta H\right)$. 

We use the following averages. $\langle\cdot\rangle$ denotes the equilibrium average
at fixed couplings, over the initial configuration, drawn from the Gibbs distribution $\propto e^{-\beta H}$,
and over the dynamical noise (the clock rings and heat-bath draws of all spins).
$\langle\cdot\rangle_\mathrm{cav}$ denotes  the same average but for the cavity system, \textit{i.e.}
the environment with spin 0 removed. An overbar
denotes the average over the coupling constants. 
Finally, ${\mathbb E}[\cdot]$ denotes the average over all the randomness of the system at hand; \textit{e.g.}, for
the SK model ${\mathbb E}[\cdot]=\overline{\langle\cdot\rangle}$.

The exact, local-in-time update of the selected spin, which will be referred to as spin number 0,
at its clock rings $t_k$, reads 
\begin{equation}\label{eq:update}
\sigma_0(t_k)=\mathrm{sgn}\big[\tanh\beta h_0(t_k^-)-v_k\big].
\end{equation}
Here  $v_k$ is a random number drawn uniformly from $(-1,1)$. The environment,
the coupling constants and the memory of the past evolution all enter through the local field $h_0(t)$.
Importantly, the evolution law is nonlinear in $h_0$. 

The basic quantity of interest is the time-dependent spin correlation function averaged over the
coupling constants,
\begin{equation}\label{eq:C}
C(t)=\overline{\langle\sigma_0(t)\sigma_0(0)\rangle}. 
\end{equation}
In the development of the theory we will also need the response function that describes the linear change
of the average value of the selected spin at time $t$ due to an external field applied to spin 0 as
$H\to H-\theta_0\sigma_0$ at $t=0^+$,
\begin{equation}\label{eq:R}
R(t)=\frac{\delta\,\overline{\langle\sigma_0(t)\rangle}}{\delta\theta_0(0)}=-\beta\dot C(t)\quad(t>0).
\end{equation}
In Eq. \eqref{eq:R} we used the fluctuation–dissipation theorem (FDT)
to relate the response and correlation functions. We define
$\Gamma=-\dot{C}(0^+) = \overline{\langle\mathrm{sech}^2\beta h_0\rangle}$. 

The distribution of the local field $h_0$ is non-Gaussian \cite{TTCSS}. However, we note that, given the value
of the selected spin, the distribution of $h_0$ is Gaussian with a mean given by the Onsager term \cite{TAP},
\begin{equation}\label{eq:law} 
P(h_0\,|\,\sigma_0)=\mathcal N\big(\sigma_0\beta J^2,\,J^2\big),
\end{equation}
where $\mathcal N(m,v)$ denotes the Gaussian distribution with mean $m$ and variance $v$.
Hence, $\Gamma$ is a Gaussian integral; at $T_c$, $\Gamma_c=0.4496$.

\section{Local field for a given selected spin history}
Let us first consider the cavity field, \textit{i.e.} the local field in the absence of the selected spin.
Removing spin 0 defines the cavity Hamiltonian $H^{(0)}$,
\begin{equation}\label{eq:Hsplit}
  H^{(0)} = H  + \sigma_0h_0.
\end{equation}
Spins evolving in the cavity system will be denoted by $\sigma_j^\mathrm{cav}$; their evolution is independent of
anything related to the selected spin. We define the cavity field $\phi(t)$ as the field acting on the selected
spin in its absence, \textit{i.e.} $\phi(t)=\sum_j J_{0j}\sigma_j^\mathrm{cav}(t)$.
In the thermodynamic limit $\phi(t)$ becomes a Gaussian process with
${\mathbb E}[\phi(t)\phi(s)]=\overline{\langle\phi(t)\phi(s)\rangle_\mathrm{cav}}=
J^2C_\mathrm{cav}(t-s)$ and $C_\mathrm{cav}(0)=1$.

We assume that at the initial time, $t=0$, the whole system is in equilibrium: $\sigma_0(0)=\pm 1$
with probability $\frac{1}{2}$ and, given $\sigma_0(0)$, the local field is distributed according to
Eq. \eqref{eq:law}. We note that the latter result can also be obtained by considering the linear 
response of the cavity to the selected spin, \textit{i.e.}, in distribution,
$h_0(0) =  \sum_j J_{0j}\sigma_j^\mathrm{cav}(0)
+ \beta\langle\phi(0)\phi(0)\rangle_\mathrm{cav}\sigma_0(0)$
(at fixed couplings). 

To obtain the field at a later time $t$ for a given path of the selected spin, $\sigma_0(t)$, we combine the
cavity field, the decay of its initial tilt, and the linear response of the cavity field to the given path,
\begin{align}\label{eq:h0t}
  & h_0(t) = \phi(t)+\beta J^2 C_\mathrm{cav}(t)\sigma_0(0)
  + \int_0^t\chi_\mathrm{cav}(t-s)\,\sigma_0(s) d s,
  \\ \label{eq:Rcav}
  & \chi_\mathrm{cav}(t) = -\beta J^2 \dot{C}_\mathrm{cav}(t).
\end{align}
To write Eq. \eqref{eq:h0t} we used the fact that for the SK model, the linear response of the
cavity field to the selected spin is exact, with higher-order terms vanishing in the thermodynamic limit.
At fixed couplings, the tilt and the kernel contain $\langle\phi(t)\phi(0)\rangle_\mathrm{cav}$,
which self-averages to $ J^2 C_\mathrm{cav}(t)$ in the thermodynamic limit.
Finally, we used FDT to write Eq. \eqref{eq:Rcav}. 

Equations (\ref{eq:h0t}--\ref{eq:Rcav}) can be condensed (integrating by parts and using $C_\mathrm{cav}(0) = 1$)
to give $h_0(t) = \phi(t) + \rho(t)$ with the dynamic Onsager reaction field $\rho(t)$ given by
\begin{equation}\label{eq:On}
  \rho(t) = \beta J^2 \left[ \sigma_0(t^-) -  \!\!\sum_{0<t_n<t}\!\!\Delta\sigma_n\,C_\mathrm{cav}(t-t_n)\right],
\end{equation}
where $\Delta\sigma_n=\pm 2$ are spin jumps at flip times $t_n$; flips occur at a subset of the clock rings $t_k$.

\section{Closing the loop: self-consistency}
The local field $h_0(t) = \phi(t) + \rho(t)$ is affine in the path of the selected spin, driven by
the cavity's randomness (its initial configuration and its noise), which is 
independent of the selected spin's noise. Thus, the actual path of the selected spin can
be inserted into the Onsager reaction field.
This step closes a causal feedback loop, solved forward in time; it is not a self-consistency condition.

To close the theory, we note that in the expression for the cavity field correlations,
$J^2 C_\mathrm{cav}(t) = \overline{\langle\phi(t)\phi(0)\rangle_\mathrm{cav}}$, 
the off-diagonal terms vanish ($J_{0j}$ are independent of the cavity system and have zero mean)
and the diagonal terms give the correlation function for any cavity spin,
\begin{equation}\label{eq:cf}
  \overline{\langle \sum_{j,k} J_{0j}J_{0k}\sigma^\mathrm{cav}_j(t) \sigma^\mathrm{cav}_k(0)\rangle_\mathrm{cav}}
  = J^2 \overline{\langle \sigma^\mathrm{cav}_k(t) \sigma^\mathrm{cav}_k(0)\rangle_\mathrm{cav}},
\end{equation}
which implies that $C_\mathrm{cav}(t)$ is the time-dependent cavity spin correlation function.
Since the cavity spins are statistically equivalent to the selected spin and since in the thermodynamic limit
the removal of spin 0 to create the cavity system does not change spin correlations, we can now
identify the correlation function of the selected spin and of the cavity spins, $C(t) = C_\mathrm{cav}(t)$.

Combining the self-consistency condition with
the exact update \eqref{eq:update}, the cavity field correlations \eqref{eq:cf} and 
the dynamic Onsager reaction field \eqref{eq:On}, and dropping the index 0, we obtain the
following closed set of equations,
\begin{align}\label{eq:updatef}
  & \sigma(t_k) =\mathrm{sgn}\big[\tanh\beta h(t_k)-v_k\big],
  \\ \label{eq:lff}
  & h(t) = \phi(t) +  \beta J^2 \left[ \sigma(t^-) -  \!\!\sum_{0<t_n<t}\!\!\Delta\sigma_n\,C(t-t_n)\right],
  \\ \label{eq:phif}
  & {\mathbb E}[\phi(t)\phi(s)]  =J^2 C(t-s),
  \\ \label{eq:Cf}
  & {\mathbb E}[\sigma(t)\sigma(s)] = C(t-s).
\end{align}
In Eq. \eqref{eq:lff}, the cavity field $\phi$ is a zero-mean stationary Gaussian process, independent
of $\sigma(0)$ and of the spin's noise, with autocorrelation function $J^2 C(t)$, Eq. \eqref{eq:phif}.

Equations (\ref{eq:updatef}--\ref{eq:Cf}) are the main result of this letter: a single unknown function, $C(t)$,
is the covariance of the cavity field and the kernel of the reaction field, and must be reproduced by the spin
it drives, Eq. \eqref{eq:Cf}.

The process described by Eqs. (\ref{eq:updatef}--\ref{eq:Cf}) starts with $\sigma(0)=\pm 1$ 
with equal probability, so that $h(0)$ is distributed according to Eq. \eqref{eq:law}.
The process describes the selected spin of a system that is in equilibrium at t = 0, and it is therefore stationary.
Its law on an interval $[0,T_w]$ depends on $C$ only on
$[0,T_w]$, so the self-consistency closes on a finite window, also at $T_c$. 

\begin{figure}[t]
\centering\includegraphics[width=\columnwidth]{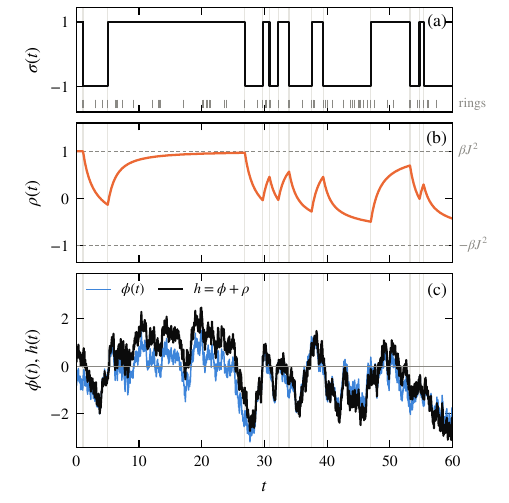}
\caption{One trajectory of the process defined by Eqs. (\ref{eq:updatef}--\ref{eq:Cf}), at $T_c$, with $C$ 
  the self-consistent solution. (a) Time dependence of the spin, with the rings of its clock (most rings do not flip it).
  (b) Time dependence of the Onsager reaction field $\rho$: it is continuous at the flips (grey lines),
  stays between $-\beta J^2$ and $\beta J^2$ (dashed lines), and relaxes toward $\beta J^2\sigma$
  as the environment repolarizes. (c) Cavity field $\phi$ and local field $h=\phi+\rho$ trajectories.}\label{fig:traj}
\end{figure}

The reaction field of a spin that has not flipped since $t = 0$ is the static Onsager field $\beta J^2 \sigma$.
At a flip, the jump of $\sigma(t^-)$ is cancelled by the new term of the sum, since $C(0) = 1$. Thus
the local field is continuous; afterwards the environment repolarizes according to $C(t)$.
Successive flips alternate in sign, so their contributions partly cancel and the most recent one dominates.
Since $C(t)$ is nonnegative and nonincreasing (by detailed balance it is completely monotone),
$|\rho|\le \beta J^2$: the reaction field never exceeds the static Onsager field. These features are illustrated in
Fig. ~\ref{fig:traj}.

Out of equilibrium, from a random start at $t=0$, the environment is initially independent of $\sigma(0)$,
so there is no initial polarization. The reasoning that leads to Eq. \eqref{eq:h0t} then gives
$h(t) = \phi(t) + J^2 \int_0^t R(t,s) \sigma(s)ds$, with ${\mathbb E}[\phi(t)\phi(s)]=J^2 C(t,s)$;
without FDT the two-time response  $R(t,s)$ is a second unknown function.
The structure of the effective process is then similar to that of
the Eissfeller--Opper process \cite{EO}.

For independent $J_{ij}$ and $J_{ji}$ the effective process becomes simpler: 
the reaction term is absent and $h(t)=\phi(t)$ \cite{CS,Grunwald}. 

\section{Critical dynamics}
To analyze dynamics upon approaching $T_c=J$, the static value,
it is convenient to introduce a function that quantifies the response to the mean local
field. Specifically, we consider adding to $h(t)$ a weak field $\theta(t)$, switched on at $t=0^+$. 
The field leaves $\phi$ and $\sigma(0)$ unchanged. It shifts the average value of the spin,
$m={\mathbb E}[\sigma]$, by
$\delta m = R*\theta$, where $*$ denotes the convolution, $(u*v)(t) = \int_0^t u(t-s)v(s) ds$. 
Since the reaction field in Eq. \eqref{eq:lff} is linear in the spin path,
the mean local field changes by $\delta{\mathbb E}[h] = \theta + J^2 R * \delta m$.
We define a 
response function $g$ through the relation between the mean spin change $\delta m$ and the mean local
field change $\delta{\mathbb E}[h]$, $\delta m= g * \delta{\mathbb E}[h]$.
Introducing the Laplace transform, $\hat{u}(z) = \int_0^\infty e^{-zt} u(t) dt$, we can write the relation between
$\hat{R}(z)$ and $\hat{g}(z)$ as follows
\begin{equation}\label{eq:Rg}
    \frac1{\hat{g}(z)}=\frac1{\hat{R}(z)}+J^2\hat{R}(z) .
\end{equation}
At $z=0$, $\hat R(0) = \int_0^\infty R(t)dt = \beta$;  this follows from FDT, since $C(0)=1$ and $C(\infty)=0$.
Eq. (15) then gives $1/\hat g(0)=(1+\beta^2J^2)/\beta$, the diagonal element of the Hessian of the
TAP free energy at $m = 0$ \cite{TAP}, \textit{i.e.}
the inverse susceptibility of a spin whose neighbors' magnetizations are held fixed; $g(t)$
is its dynamical counterpart.

Eq. \eqref{eq:Rg} can be solved for $\hat{R}$,
\begin{equation}\label{eq:Rg2}
  \hat R(z)=\frac{1-\sqrt{1-4J^2\hat g^2(z)}}{2J^2\hat g(z)}
  \end{equation}
The square root in Eq.~\eqref{eq:Rg2} has a branch point $z_*$, where $2J\hat g(z_*)=1$.
Above $T_c$, it lies on the negative real axis, and $-z_*$ is the slowest relaxation rate of the response and
correlation functions, $R(t)$ and $C(t)$. 
We define the mean time of the new response function, $g(t)$, as $\tau_g=\int t\,g\,d t/\!\int g\,d t$. 
Then, near $z=0$, $\hat g(z)\simeq\hat g(0)(1-\tau_gz)$. Expanding $1-4J^2\hat g^2(z)$ to first order in $z$
in Eq. \eqref{eq:Rg2} allows us to identify the branch point,
\begin{equation}\label{eq:zstar}
z_*\simeq-\frac{(1-\beta^2J^2)^2}{8\beta^2J^2\,\tau_g} .
\end{equation} 
Eq. \eqref{eq:zstar} shows that $z_*<0$ above $T_c$, and $z_*\to 0$ at $T_c=J$,
provided $\tau_g$ stays bounded as $T\to T_c$. The slowest rate then vanishes as $(T-T_c)^2$ and the relaxation
time diverges as $(T-T_c)^{-2}$.

Next, we analyze the dynamics at $T_c$. We assume that at long times the correlation function decays
as a power law, $C(t) \simeq A t^{-a}$ with $0<a<1$, and we check what power law is allowed by the theory.
We define $\hat y(z) = z \hat C(z)$ and note that at $T_c$ FDT gives $J\hat R(z) = 1 - \hat y(z)$.
Then, from Eq. \eqref{eq:Rg} we get
\begin{equation}\label{eq:gy}
  2J\hat g = (1-\hat y) / (1-\hat y + \hat y^2/2)
\end{equation}
(we omit the arguments to lighten the notation). For small $\hat y$ we get
$1-2J\hat g \simeq \hat y^2/2$. The assumed power-law decay of $C(t)$ leads to $\hat y \propto z^a$ and
thus $1-2J\hat g \propto z^{2a}$. This behavior has to match the previously assumed form of the
small-$z$ expansion of $\hat g$, which at $T_c$ reads $\hat g \simeq (1/2J)(1 - \tau_g z)$.

We see that for $a>1/2$ (and also any $C(t)$ with a finite integral) the two expressions can match
only if $\tau_g =0$, which is impossible for a positive $g(t)$.

For $a < 1/2$, $1 - 2J\hat g(z)\propto  z^{2a}$ would require a tail $g(t) \propto t^{-1-2a}$.
However, the long-time part of $g(t)$, when expanded in terms of the long-time parts of $C(t)$ and $R(t)$,
starts with terms of the type $C^2(t)R(t) \propto t^{-1-3a}$:
terms without a response function vanish by causality, a single $R(t)$ is the reaction chain already summed
in Eq. \eqref{eq:Rg}, and the symmetry $\sigma\to -\sigma$, $\phi\to -\phi$ excludes terms $R(t)C(t)$.
Terms  $C^2(t)R(t) $ contribute to $1 - 2J\hat g(z)$ at order $z^{3a}$, not $z^{2a}$.

\begin{figure}[t]
\centering\includegraphics[width=\columnwidth]{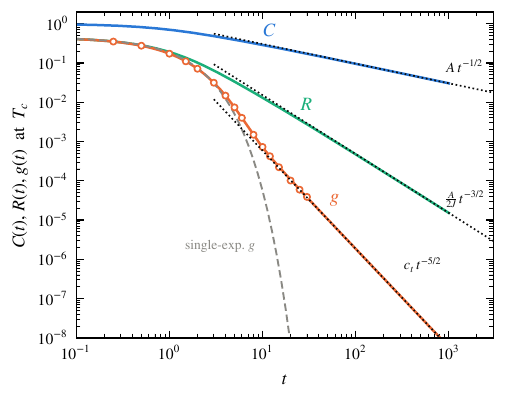}
\caption{Time dependence of $C(t)$, $R(t)$ and the response to the mean local field $g(t)$, at $T_c$.
  Lines: the self-consistent solution. Dotted: the exact asymptotes $At^{-1/2}$, $(A/2J)t^{-3/2}$ and
  $c_t t^{-5/2}$, Eqs. \eqref{eq:critical2}, \eqref{eq:R} and (\ref{eq:constsgtail}--\ref{eq:constants}),
  with $A = 0.964$ and $c_t = 0.186$; the extracted $g(t)$ at $t = 20$--$30$ agrees well with $c_t t^{-5/2}$.
  Dashed: the single-exponential g of the linear Mori--Zwanzig closure \cite{Szamelsg2}.
  Circles: $g(t)$ extracted from the single-spin process driven by the self-consistent $C(t)$.
  The slow decay of $C(t)$ and $R(t)$ is generated by the square root in Eq. \eqref{eq:Rg2} from a short-ranged
  $g(t)$.}\label{fig:gRC}
\end{figure}

The above arguments imply that $a=1/2$, with finite $\tau_g$, and thus
\begin{equation}\label{eq:critical1}
  \hat R\simeq\frac{1-\sqrt{2\tau_gz}}{J}
\end{equation}
\begin{equation}\label{eq:critical2}
  C(t)\simeq A\,t^{-1/2},\qquad A=\sqrt{\frac{2\tau_g}{\pi}} .
\end{equation}
Inserting $\hat C=A\sqrt\pi\,z^{-1/2}+c_1+\dots$ into Eq. \eqref{eq:gy} gives
\begin{equation}\label{eq:gzsmall}
  2J\hat g=\frac{1-\hat y}{1-\hat y+\hat y^2/2}=1-\tau_gz+\gamma z^{3/2}+\dots,
\end{equation}
\begin{equation}\label{eq:constsgtail}
  \gamma=-A\sqrt\pi\,(c_1+\tau_g),\qquad g(t)\simeq c_t\,t^{-5/2},
\end{equation}
\begin{equation}\label{eq:constants}
c_t=-\frac{3A}{8J}\big(c_1+\tau_g\big),\qquad c_1=\int_0^\infty\!\big[C-At^{-1/2}\big] dt.
\end{equation}
Given the $t^{-1/2}$ tail of $C(t)$, the structure of Eq. \eqref{eq:Rg}
allows no non-analytic term of $\hat g(z)$ below $z^{3/2}$,
and fixes its coefficient in terms of $A$, $c_1$ and $\tau_g$; however, it does not require this term to be present. 
A rational $\hat g(z)$ (\textit{i.e.} a finite sum of exponentials in the time domain) has $\gamma=0$,
and then $c_1=-\tau_g$ exactly. For the numerical solution, the $C^2(t)R(t)$ terms
discussed above make it positive: $c_1=-1.97<-\tau_g=-1.46$, so that $\gamma=0.88$ and $c_t=0.19$
(Fig.~\ref{fig:gRC}).
The relation \eqref{eq:gy} fixes the form of the tail, not the value of the prefactor, $A$.

The exponent $1/2$ thus follows from Eq. \eqref{eq:Rg}, FDT and two properties of $g(t)$,
its positivity and the $C^2(t)R(t)$ structure of its long-time part, independently of the details of $g$.
Both properties are borne out by the solution: the extracted $g(t)$ is positive and decays as $t^{-5/2}$
(Fig. \ref{fig:gRC}).
In contrast, $A$ depends on the details of $g(t)$: it is fixed by the mean time of $g(t)$,
which none of the exact relations used above determines.
It follows only from the self-consistency, \textit{i.e.} from the numerical solution described below,
and it depends on the dynamical rule \cite{BC}.

Finally, the $(T-T_c)^{-2}$ law of Eq. \eqref{eq:zstar} requires $\tau_g$ to stay bounded as $T\to T_c$.
Above $T_c$, Eq. \eqref{eq:Rg} at small $z$ gives the exact relation
$\tau_g=\tau_{\rm int}(1-\beta^2J^2)/(1+\beta^2J^2)$, with $\tau_{\rm int}=\int_0^\infty C\,dt$;
for the solution it gives $\tau_g=1.01$, $1.03$ and $1.10$ at $2T_c$, $1.5T_c$ and $1.25T_c$,
against $1.46$ at $T_c$.

\section{Solving the closed process}
We use two different strategies to solve Eqs. (\ref{eq:updatef}--\ref{eq:Cf}) above and at $T_c$:
we iterate in $C(t)$ above $T_c$ and in $g(t)$ at $T_c$. 

We represent $C(t)$ as a positive mixture of exponentials $C(t)=\sum_k a_k e^{-\nu_kt}$
($C(t)$ is completely monotone), with the exact $C(0)=1$, $\dot{C}(0^+)=-\Gamma$ and
$\ddot{C}(0^+)=\Gamma$.  Eq. \eqref{eq:lff} is then a sum of exponentially decaying variables and
$\phi(t)$ is a sum of independent Ornstein–Uhlenbeck processes.
Together with $\sigma$ these variables form a Markov process, which starts in its stationary distribution
(whose one-time marginal is Eq. \eqref{eq:law}) and is reversible. The process is simulated exactly,
with updates only at the clock rings.

Started in this equilibrium distribution,
the process is stationary for any input, not only at the fixed point, and its output obeys
FDT, $R_\mathrm{out}(t) = -\beta \dot C_\mathrm{out}(t)$. Thus $C_\mathrm{out}$
can be estimated from all time origins.

Above $T_c$,
plain iteration in $C(t)$ converges. We drive the process defined by Eqs. (\ref{eq:updatef}--\ref{eq:phif})
by $C=C_\mathrm{in}$ and evaluate $C_\mathrm{out}$ from the spin correlation function,
${\mathbb E}[\sigma(t+\tau)\sigma(t)]$. This defines a map $C_\mathrm{in} \to C_\mathrm{out}$.
The theoretical prediction for the spin correlations is the fixed point of this map, which is found by iteration.

At $T=T_c$, we use the numerical strategy suggested by Eq. \eqref{eq:critical2}.
We note that 
the amplitude $A$ of the critical decay is a property of
the asymptotic tail of $C(t)$,
which is approached slowly: for the solution, $C(t)\sqrt{t} = 0.912, 0.957$ and $0.963$ at
$t = 10, 100$ and $1000$. In contrast, in terms of $g(t)$,
$A$ is given by the first moment of a short-ranged function (Fig. \ref{fig:gRC}). 
Thus, at $T_c$ we iterate in $g(t)$.
We parametrize it as a positive mixture of exponentials plus a band of slow rates,
$g(t)=\sum_k b_ke^{-\mu_kt}+c\!\int_0^{\mu_c}\!\mu^{3/2}e^{-\mu t} d\mu$,
with $b_k, c \ge 0$; the integral term produces the $t^{-5/2}$ tail indicated in Eq. \eqref{eq:constsgtail}
beyond $t\approx 1/\mu_c$.
The model is constrained by the exact values $g(0^+)=\beta\Gamma$, $\dot g(0^+)$, $\ddot g(0^+)$,
and the sum rule $\hat g(0)=\beta/(1+\beta^2J^2)$.

One iteration takes four steps: (1) $C_{\rm in}(t)$ is computed from $g(t)$ using Eq. \eqref{eq:Rg2}, 
FDT and a numerical Laplace inversion, and represented by a mixture on $0\le t\le 160$, where
the process is run (its tail beyond is not needed, see ``Closing the loop''); (2) the process
is run to give $C_{\rm out}(t)$; (3) Eq.~(15) gives $g_{\rm out}(t)$ on the interval
$0\le t\le60$; (4) the model is refitted to $g_{\rm out}(t)$.

Starting from the single-exponential $g(t)$ of the linear Mori--Zwanzig closure,
$\tau_g$ reaches $1.43$--$1.45$ within three to
four iterations. The solution is the mean of twelve converged models
(band edges $\mu_c$ = 0.3 and 0.6, six independent seeds each):
\begin{equation}\label{eq:Tcresult}
\tau_g(T_c)=1.46\pm0.02,\qquad A=0.964 \pm 0.007 .
\end{equation}
About 83\% of the first moment of $g(t)$, and hence of $\tau_g$, comes from $t\le 10$, where $g(t)$
is obtained directly from the single-spin process; the rest comes from the tail, whose weight is fixed
by the sum rule and whose form is the $t^{-5/2}$ law of Eq. \eqref{eq:constsgtail}, seen in the extracted
$g(t)$ in Fig. \ref{fig:gRC}. The quoted error of $\tau_g$ is set by the uncertainty in the form of the tail
(subleading corrections to $t^{-5/2}$); the statistical error is 0.003.
As an independent check, forward substitution \cite{EO}, which builds $C(t)$ lag by lag without iteration
and without any model of $g(t)$, agrees with the solution within one standard error at all times and gives
$A = 0.963 \pm 0.004$.

The earlier theories can be interpreted as approximations of $g(t)$ as defined by Eq.~\eqref{eq:Rg}. 
The single-exponential $g(t)$
of the linear Mori--Zwanzig closure (the zeroth-order theory of Refs.~\cite{Szamelsg1,Szamelsg2})
gives $\tau_g=1.11$ and $A=0.841$. The two-exponential $g(t)$ of the first-order theory~\cite{Szamelsg2},
which reproduces the first three derivatives of $C(t)$ at $t = 0^+$ (equivalently $g(0^+), \dot g(0^+)$ and
$\ddot g(0^+)$) and the sum rule exactly, has $\tau_g=1.21$ and
$A=0.878$. Sommers' expansion~\cite{Sommers} corresponds to $\tau_g=1$ and $A=0.798$.
We should emphasize that all these theories predict the correct $\propto t^{-1/2}$ decay of spin
correlations, since their $\tau_g$ is finite (the exponential ones even without any algebraic tail in $g(t)$).
They all underestimate $A$ because their $g(t)$ decays too fast: exact short-time behavior does not fix $\tau_g$.

\begin{figure}[t]
\centering\includegraphics[width=\columnwidth]{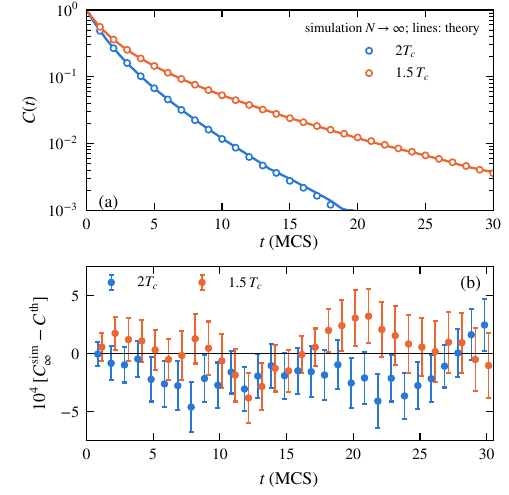}
\caption{Comparison of computer simulation and theoretical results above $T_c$.
  (a) $C(t)$ at $2T_c$ and $1.5\,T_c$. Circles: simulations extrapolated to $N\to\infty$.
  Lines: the self-consistent theory (plain iteration in $C(t)$).
  (b) Differences, in units of $10^{-4}$, between simulation and theory,
  with combined errors.}\label{fig:hiT}
\end{figure}

\section{Computer simulation tests of the theory}
To test the theory we ran computer simulations of Glauber dynamics of the SK model.
Individual spins to be updated were selected randomly (random-sequential updates).
Local fields were stored and updated at spin flips \cite{MY}. 
One Monte Carlo step per spin (MCS), \text{i.e.} $N$ updates,
is used as a unit of time. In the $N\to \infty$ limit we get the continuous-time
dynamics of Eq. \eqref{eq:update}. 
$C_N(t)$ is the autocorrelation at size $N$, averaged over spins, time origins and coupling samples;
$C_\infty(t)$ is its extrapolation to $N\to\infty$.

Above $T_c$, at $2T_c$ and $1.5T_c$, we studied systems consisting of  $N=256$--$8192$ spins, with $32$ to $4$
coupling samples. The simulations were run for $5000$--$6000$ MCS each, with the first $500$ MCS discarded.
Finite-size corrections are $O(1/N)$ and are already small at $N=8192$;
$C_\infty(t)$ is a linear extrapolation in $1/N$.

\begin{figure*}[t]
\centering\includegraphics[width=\textwidth]{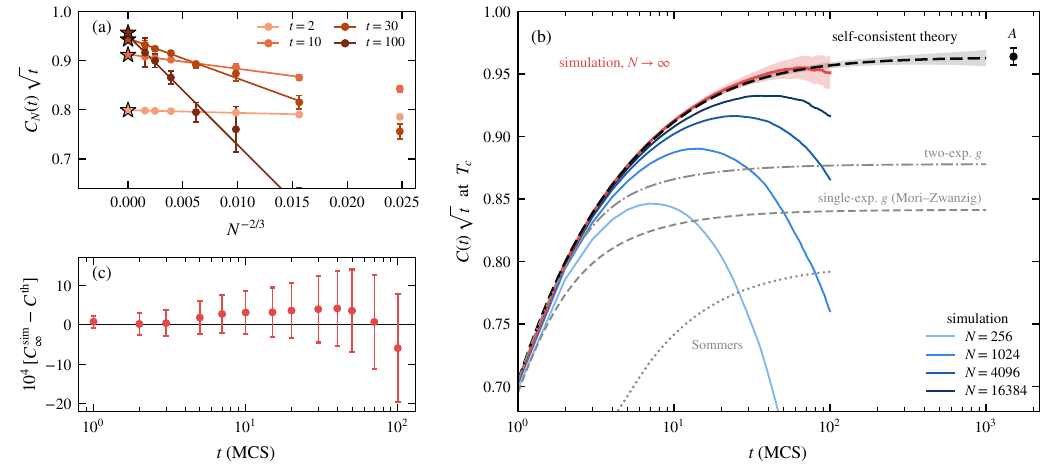}
\caption{Spin correlations at $T_c$. (a) $C_N(t)\sqrt{t}$ against $N^{-2/3}$ at four time lags,
for all seven sizes. Lines: linear fits to $N\ge 512$; stars: theoretical predictions. (b) $C(t)\sqrt{t}$:
simulations at four sizes, their extrapolation to $N\to\infty$ (band: $\pm 1$ standard
deviation), and the self-consistent theory with its error band, continued to $t=1000$. The
point at the right is the asymptote $A$, Eq.~\eqref{eq:Tcresult}. Grey: the single-exponential $g(t)$ (linear
Mori--Zwanzig closure), the two-exponential $g(t)$ (first-order theory) and Sommers' expansion.
(c) Difference between the extrapolated simulations and the theory, in units of $10^{-4}$,
with combined errors. The error of the theory, $1$--$2.6\times10^{-4}$, combines the scatter
of the twelve models, the two band edges and a systematic error of the refit.}\label{fig:Tc}
\end{figure*}

The comparison between the simulation and theory results is shown in Fig. \ref{fig:hiT}.
They differ by at most $1.8$ standard deviations at $1.5\,T_c$ and $2.2$ standard deviations at $2T_c$.
The simulation errors are $0.4$--$2\times10^{-4}$.

To test the theory further, we used the simulated spin correlations $C_\infty(t)$ at $1.5\,T_c$ as input of the
map. It was returned within $1.5$ standard deviations, thus confirming that it is a fixed point of the theory.

At $T_c$, we studied $N=256$--$16384$ spins,
with $64$ to $9$ coupling samples. The simulations were run for $8000$--$12000$ MCS each.
The discarded time grew with the system size as $N^{2/3}$, up to $5500$ MCS. 

At $T_c$, the finite-size effects require more care.
The critical window of the SK model has width $N^{-1/3}$ in $T- T_c$ \cite{BC,BZ}.
By Eq. \eqref{eq:zstar}, the relaxation time at $T_c$ is then cut off at $\tau_N  \propto N^{2/3}$, and
\begin{equation}\label{fss}
  C_N(t) = t^{-1/2} \Phi(t/N^{2/3}),
\end{equation}
the scaling found by Billoire and Campbell \cite{BC}.
For $t\ll N^{2/3}$, a regular $\Phi(x) \simeq A(1 - cx)$ gives $C_N(t) - C_\infty(t) \simeq -Ac t^{1/2}N^{-2/3}$.
The data confirm both powers: at $t = 2$--$10$, where $t/N^{2/3} \le 0.25$ for all sizes, the correction scales
as $t^{0.53\pm 0.06}N^{-0.71\pm 0.06}$. We therefore extrapolate $C_N(t)$ linearly in $N^{-2/3}$ at each time lag,
as shown in Fig. \ref{fig:Tc}a.

The extrapolated spin correlation function is compared with the theoretical result in Fig. \ref{fig:Tc}b.
They agree within one standard deviation, Fig. \ref{fig:Tc}c; no systematic difference is observed. The three earlier
theories are excluded by far more than five standard deviations.

\section{Discussion and outlook}
We have analyzed Glauber dynamics of the Ising spin SK model at zero external field, above and at the
spin glass transition temperature. We reduced the many-body dynamics of interacting spins to a closed,
single-spin process, Eqs. (\ref{eq:updatef}--\ref{eq:Cf}), with a single unknown function $C(t)$.
The spin updates are exact, local in time; the memory resides in the local field. The spin correlation
function $C(t)$ encodes both the fluctuations of the local field and the kernel of the reaction field. 
The solution agrees with simulations extrapolated to $N\to\infty$ to within a few $10^{-4}$, above
as well as at $T_c$. 

The statement that the theory is exact for $N\to\infty$ rests on two standard ingredients of the dynamic
cavity method: the linear response of the cavity to the selected spin, and the Gaussian statistics of the cavity field.
Both have been proved in related settings \cite{Grunwald,DGPZ}, but not for Glauber dynamics of the SK model. 

Our approach can be extended in several directions. Out of equilibrium, time-translation
invariance is lost and the response $R(t,s)$ becomes a second unknown function; it can be
obtained from the same single-spin process through the Novikov identity,
$\mathbb{E}[\phi(s)\sigma(t)]=J^2\int_0^t C(s,u)R(t,u)\,du$, as in the work of Eissfeller
and Opper \cite{EO}. Below $T_c$, $C(t)$ tends to the Edwards--Anderson order parameter $q>0$, the
cavity field acquires a frozen part of variance $J^2q$ [3], and after a quench the dynamics
ages \cite{CK2}. These extensions are left for future work.

\acknowledgments
I thank F. Zamponi and P. Charbonneau for discussions on using LLMs in research.
The theory was derived with the help of Claude (Anthropic), used to analyze many different proposals.
I also used Claude (Anthropic) to perform numerical analysis of the theory, run the simulations,
prepare the figures and check the grammar.
This research was supported in part by NSF Grant~PHY 2309135 to the Kavli Institute for Theoretical Physics (KITP).
The hospitality of the Yukawa Institute for Theoretical Physics, where this work was finished, is gratefully
acknowledged.

\end{document}